\documentclass[conference]{IEEEtran} 

\IEEEoverridecommandlockouts
\IEEEpubid{\begin{minipage}[t]{\textwidth}\ \\[11pt]
        \centering\normalsize{978-1-7281-5842-6/19/\$31.00~\copyright~2019 IEEE}
\end{minipage}}

\usepackage[dvips]{graphicx}
\usepackage{subcaption}
\usepackage{epsf}
\usepackage{epsfig}
\usepackage[noadjust]{cite}
\usepackage{amsmath}
\usepackage{url}
\usepackage[export]{adjustbox}
\usepackage{soul}
\usepackage{algorithm}
\usepackage[noend]{algpseudocode}
\usepackage{color}
\usepackage{tree-dvips}
\usepackage{float}
\usepackage{multirow}
\usepackage{comment}
\usepackage{soul}
\usepackage{booktabs}
\usepackage[dvipsnames]{xcolor}
\usepackage{siunitx}
\usepackage{footnote}
\usepackage{threeparttable}
\usepackage{pifont}
\usepackage[T1]{fontenc}
\usepackage[algo2e]{algorithm2e}

\newcommand{\cmark}{\ding{51}}%
\newcommand{\xmark}{\ding{55}}%

\def\BibTeX{{\rm B\kern-.05em{\sc i\kern-.025em b}\kern-.08em
    T\kern-.1667em\lower.7ex\hbox{E}\kern-.125emX}}

\usepackage{eso-pic}
\newcommand\AtPageUpperMyright[1]{\AtPageUpperLeft{%
 \put(\LenToUnit{0.5\paperwidth},\LenToUnit{-1cm}){%
     \parbox{0.5\textwidth}{\raggedleft\fontsize{9}{11}\selectfont #1}}%
 }}%
\newcommand{\conf}[1]{%
\AddToShipoutPictureBG*{%
\AtPageUpperMyright{#1}
}
}

\newsavebox{\mybox}
\newlength{\mydepth}
\newlength{\myheight}

{\begin{lrbox}{\mybox}\begin{minipage}{\textwidth}}%
{\end{minipage}\end{lrbox}%
  \settodepth{\mydepth}{\usebox{\mybox}}%
  \settoheight{\myheight}{\usebox{\mybox}}%
  \addtolength{\myheight}{\mydepth}%
  \noindent\makebox[0pt]{\hspace{-20pt}\rule[-\mydepth]{1pt}{\myheight}}%
  \usebox{\mybox}}
  
\begin{document}

\onecolumn 

{\LARGE IEEE Copyright Notice} \\

\copyright 2019 IEEE. Personal use of this material is permitted. Permission from IEEE must be obtained for all other uses, in any current or future media, including reprinting/republishing this material for advertising or promotional purposes, creating
new collective works, for resale or redistribution to servers or lists, or reuse of any copyrighted component of this work in other works. \\

{\large Accepted to be Published in: Proceedings of the 2019 22nd International Conference on Computer and Information Technology (ICCIT), 18-20 December 2019, Dhaka, Bangladesh.}

\twocolumn

\title{Socio-network Analysis of RTL Designs for Hardware Trojan Localization} 





\author{
    \IEEEauthorblockN{Sheikh Ariful Islam\IEEEauthorrefmark{1}, Farha Islam Mime\IEEEauthorrefmark{2}, S M Asaduzzaman\IEEEauthorrefmark{3}, Farzana Islam\IEEEauthorrefmark{4}}
    \IEEEauthorblockA{\IEEEauthorrefmark{1}University of South Florida || Email: sheikhariful@mail.usf.edu}
    \IEEEauthorblockA{\IEEEauthorrefmark{2}University of Information Technology and Sciences || Email: farhaislam.eee@gmail.com}
    \IEEEauthorblockA{\IEEEauthorrefmark{3}ABB Limited || Email: asaduzzaman.tito@gmail.com}
    \IEEEauthorblockA{\IEEEauthorrefmark{4}Khulna University || Email: farzanaemu98@gmail.com}
}

\maketitle
\conf{2019 22nd International Conference on Computer and Information Technology (ICCIT), 18-20 December 2019}

\begin{abstract}
The recent surge in hardware security is significant due to offshoring the proprietary Intellectual property (IP). One distinct dimension of the disruptive threat is  malicious logic insertion, also known as Hardware Trojan (HT). HT subverts the normal operations of a device stealthily. The diversity in HTs activation mechanisms and their location in  design brings no catch-all detection techniques.  In this paper, we propose to leverage principle features of social network analysis to security analysis of Register Transfer Level (RTL) designs against HT. The approach is based on investigating design properties, and it extends the current detection techniques. In particular, we perform both node- and graph-level analysis to determine the direct and indirect interactions between nets in a design. This technique helps not only in finding vulnerable  nets that can act as HT triggering signals but also their interactions to influence a particular net to act as HT payload signal. We experiment the technique on 420 combinational HT instances, and on average, we can detect both triggering and payload signals with accuracy up to  97.37\%.

\end{abstract}

\section{Introduction}
\label{sec:Intro}

The sheer complexity of modern electronic systems and their distributed supply chain spread across the globe are growing concerns for establishing trust in Integrated Circuits (IC) and systems. The untrusted third-parties involved in fabrication, assembly, packaging, testing, and validation can make malicious modifications to the ICs to undermine the security of the system. Such intentional changes, known as Hardware Trojan (HT), are typically performed without designer knowledge, and it works as a backdoor to achieve undesired behavior. The potential threats of HT are often aligned with the attackers' objectives that include but not limited to the leaking secret information (e.g., key), incorrect functionality, early failure of the device, etc. Vulnerabilities of the design are often stipulated during pre- and/or post-silicon where a small HT circuit (3-5x smaller than regular design) observes certain events and becomes active at an infrequent time when the device is in the field. The part of the design showing limited controllability and observability can provide the required stealthy behavior of HT. Furthermore, the spectrum of HTs depending on their physical, activation, and action characteristics \cite{5406669} make the current detection approaches non-trivial \cite{Xiao:2016:HTL:2948199.2906147}.

The construction of concurrent HTs is concerned with the nets that switch rarely within a design. Rare triggering nets directly influence the location and insertion phase (gate-level or layout-level). As more and more Commercial Off The Shelf (COTS) components are integrated within the system without security verification but to meet time-to-market demands, the attack surface becomes more dynamic. Hence, traditional detection approaches need to be transformed to account for newly published HTs.

During pre-silicon, a defender analyzes the security by performing logic testing of the design against HT vulnerability. Although logic testing is independent of the process variations, it is limited by the simulation cost, which depends on design size and the modeling of the simulator. Such a simulation-based technique involves generating random test vector \cite{Huang:2016:MST:2976749.2978396} or high-level statistical modeling \cite{8607170,8885006,8357321} of the design to propagate the test vector through the design. Regardless of the vector generation technique, the analysis of the switching activity of all nodes in a design can identify the nets with the lowest activity. Based on the study, a designer can generate region-specific test vector to improve the controllability and observability of these nets \cite{cryptoeprint:2015:1252,4559047}.  

On the contrary, Side-Channel Analysis (SCA)  is performed to detect minute variations in design parameters (area, power, and delay) due to the presence of HT \cite{4223234} during post-silicon. SCA treats the design as a whole and collects the signature and compare it with HT-free golden design. The strength of SCA is challenging due to the decreasing resolution of HT size and increasing process variations in sub-nm design \cite{4559049,8839352}. Many of works for HT detection and prevention require pre-characterization of HT-free design which may not be available sometime. For example, IP vendors provide limited information about the internal construction of IP in their published datasets. Furthermore, the limitations of logic/functional testing in generating the same test vectors for both attacker and defender are nearly infeasible due to ample logical space, and the mechanism of reliability failures may even mask HT presence. Hence, a common assumption regarding golden design makes HT detection more challenging and intractable as to learn the perturbed nets by an attacker and identify them through SCA.

In this paper, we utilize the concept of the social network to delineate the relation between design properties and  locate HT within a design.  As a first step, we perform a formal treatment of circuit edge (nets) as it is the heart of HT triggering. We use essential attributes of edges since adding or removing edges affect the neighborhood of other edges as well as design parameters. Then, identified characteristics of edges are ranked to understand the potential perturbations over edges. It also helps to gain insight into an evolving network (triggering signals) and how it influences other edges (payload). As there can exist multiple of such network, we analyze each of this network cluster who can perform collusion by adding or removing edges strategically. Then, based on both individual edge and group network, compact test vector can be generated for inter-cluster. These test patterns would be completely localized without resorting to logic simulation {\em a priori} and magnify the probable HT location (if any). The novelty and contributions of the proposed approach are:
\begin{itemize}
    \item there is no assumption in underlying design (HT-free or HT-affected), hence the technique is uniform.
    \item a comprehensive framework for combinational HT detection based on individual (node) and cluster (network) analysis to identify rare triggering nets.
    \item a localized test pattern generation technique for  neighborhood isomorphism.
\end{itemize}

The rest of the paper is organized as follows. Section \ref{sec:back} provides an overview of related works on HT localization based on local vs. global excitation. Section \ref{sec:proposed} describes attack model, related definitions and bottom-up method to detect HT. Section \ref{sec:result} presents the detection capability of the proposed technique without any knowledge of golden design. Section \ref{sec:conclude} draws the conclusion followed by future work.

\section{Background and Related Work}
\label{sec:back}

In this section, we present the existing HT localization techniques during both pre- and post-silicon. As our focus in this paper is the detection of combinational HT, we review the works that apply heuristics to particular nodes to act them as HT triggering signal and payload. Many of these works rely on extracting (rare) transition activity before they use a specific HT detection methodology. Hence, we broadly classify simulation technique under two conditions: (a) if individual edge \footnote{Unless stated elsewhere in the paper, edge and net are used interchangeably.} attributes are considered, and (b) if combined properties of only a subset of edges are considered without checking the individual net.

Zhou \emph{et al.} \cite{6971844} proposed transition probability enhancement of the nets far from primary inputs by using two input MUX. The approach is practical for gate-level netlist, but for low-level (e.g., RT-level), the technique is not scalable. Zhou \emph{et al.} \cite{8368626} approached the combinational HT by fault simulation (stuck-at). The authors have considered only one type fault during excitation of a net, however, an attacker can enumerate multiple nets with different fault model. 

For the sub-network based approach, Banga and Hsiao \cite{4559047} presented a region of interest around a gate that contains flip-flops. The technique was successful in detecting sequential HT, however, the proposed approach is applicable for combinational HT. Koushanfar and Mirhoseini \cite{5657256} applied gate profiling technique to detect side-channel parameters. As the HT detection problem is NP-hard, the impact of process variations can be significant on the individual gate, hence can mask the HT presence. Wei and Potkonjak \cite{5653770} presented gate-level characterization approach where the authors partitioned the design based on input vector control and profile region-based current. The technique, however, fails to detect HT if they are located in multiple regions. Banga and Hsiao \cite{5513114} presented miter circuit to detect equivalence between golden and HT-infected design. The circuit would produce {\tt UNSAT} if there is any HT. 

Unlike previous techniques, our technique (a) does not consider any HT excitation by expensive simulation, instead, applies network parameters, and (b) is scalable for both combinational and sequential HT as it classifies most influential nodes for HT triggering without any heuristics.



\section{Proposed Approach}
\label{sec:proposed}

\subsection{Threat Model}
\label{sec:threat}

We assume the potential inclusion of out-of-spec components into the legacy design can happen during pre- and post-silicon. In pre-silicon, third-party IP vendors can modify parts of the design and sell to the particular IP buyer where the buyer may be fabless design house. Similarly, a system integrator can act as an insider attacker where he has access to design internals and can subvert any IP cores included in System-on-Chip (SoC). Even the IP core is encrypted, as a valid and regular user, the integrator would have the correct key to unlock the core. During post-silicon, an attacker in the foundry would have access to GDSII file which he can reverse engineer to insert HT by changing process parameters (e.g., lithographic mask).

\subsection{Problem Statement}
\label{sec:problem}

A combinational circuit can be modeled as Directed Acyclic Graph (DAG), G (V, E) where V(G) denotes the set of nodes and E(G) is the directed edge set. The problem involves identifying the set of edges' properties, which are  potential candidates for HT triggering and payload. To do so, we convert G to the edge-labeled graph, G'(E, V) where, E(G') denotes nodes (conversely, edges in G) and V(G') indicates the edge set (conversely, nodes in G). In our case, to ensure localization of possible HT, we require to identify at least $k$ candidate edges in G with high-confidence using the properties defined in Section \ref{sec:def}.

\subsection{Definitions and Terminology}
\label{sec:def}

{\textbf{Centrality, C}: returns the number of adjacent nodes to a particular actor and includes both in- and out-degree of a node. Hence, nodes having more impact on HT should have lower centrality and follow an inverse power-law distribution. If A$_{ij}$ is the adjacency matrix in G', we can compute C of a node as follows:
\begin{equation}
    C_{i} = \frac{\sum_{j} A_{ij}}{|A_{ij}|}
     \end{equation}
where $|A_{ij}|$ denotes the total locations where edge exists (i.e. A$_{ij}$=1).
}

{\textbf{Closeness Centrality, CC}: determines the shortest path length from a node to all other nodes. The nodes in G' showing the lowest CC (maximum average path length) should be identified as they can manifest themselves as low controllable and observable nodes \cite{dupuis2013identification}. If the number of edges traversed between two nodes $i$ and $j$ in the shortest path is s$_{ij}$, CC can be computed as follows:
\begin{equation}
    CC_{i} = \frac{\sum_{j} s_{ij}}{|s_{ij}|}
     \end{equation}
where $|s_{ij}|$ denotes the total number of mutually exclusive shortest paths.
}

{\textbf{Betweeness Centrality, BC}: acts as a bridge for shortest paths between two nodes. In general, if  the BC value of a node in G' is higher, the higher is  the chance the node would act HT payload as the node is reachable from other two nodes showing lowest CC. Conversely, a new node can be added to G' to perform collusion between two sub-groups in a design. We can calculate BC of a node as follows \cite{7842049}: 
\begin{equation}
    BC_{i} = \frac{1}{N(N-1)} \sum_{j=1}^{j=N} \sum_{k=1,k \neq j}^{k=N} I_{p_{j,k}} (n_{i})
     \end{equation}

where the value of $I_{p_{j,k}}$($n_{i}$) is `1' if shortest path exists between $j$ and $k$ and it passes through $i$.
}

{\textbf{Eigen Vector Centrality, EVC}: computes the centrality of a node by accounting centrality of nodes in its neighborhood. Once we calculate the most influential nodes in G' from C and CC metrics, EVC can estimate the relative importance of another node(s) in the periphery of the influential node. Similar to BC, node(s) having highest EVC  can work as payload, and we can compute EVC by taking principle eigenvalue ($\lambda$) of the adjacency matrix between HT triggering nodes and payload node.}

{\textbf{PageRank, PR}: estimates the probability that a node can be inserted in G' based on the edge information of neighborhood nodes. As the location of the future edge(s) is essential, and attributes of nodes are entirely available in  design \cite{YAO201682}, we apply intimacy between neighbors to describe the likelihood of edge.   }

{\textbf{k-Cores, k$_{c}$}: is a connected sub-graph of the network where each node in the sub-graph has at least degree of k. Once we compute C, CC, and BC,  we can find the ``degeneracy'' of the HT triggering cluster (highest value of k).}

{\textbf{Density, D}: denotes the connectivity information of a sub-graph or graph. For strongly connected graph/sub-graph, D is close to 1, and an attacker would avoid it as inserting HT in this graph/sub-graph would influence more in changing design parameters. If the information about the centrality and its derivatives are not available, we can  start with K$_{c}$=1 to get the density of each sub-network(s) in the graph and rank them from lower to higher. The cluster with the lowest rank can be chosen to embed HT.}

\begin{table}
\begin{center}
\caption{Impact of socio-network definitions on HT parameters analysis }
\label{tab:sign_example}
{
\begin{tabular}{|c| c| c| c| c| c|}
\hline
Analysis & Definitions  & \multicolumn{2}{c|}{ Impact} \\
\cline{3-4}
 Type & & Triggering & Payload \\ \midrule
 \multirow{5}{*}{Node-level} & C & \cmark & \xmark\\
 & CC & \cmark & \xmark\\ 
 & BC & \xmark &  \cmark \\ 
 & EVC & \xmark& \cmark \\ 
 & PR & \xmark & \cmark \\ \midrule
\multirow{2}{*}{Graph-level} & k-cores & \cmark & \cmark \\ 
 & Density & \cmark & \cmark\\ \midrule
\end{tabular}
}
\end{center}
\vspace{-5ex}
\end{table}

\subsection{Motivational Example}
\label{sec:motivational}

We address the problem of HT localization using a bottom-up technique, which is a fine-grained approach. We start with removing the Primary Inputs (P.I.), and Primary Outputs (P.O.) from DAG of design as inserting HT at the periphery of a design can be easily detected. Then, we convert the node-labeled DAG (G) into edge-labeled DAG (G'). We concentrate on two crucial analysis (node- and graph-level). Node-level analysis is mainly concerned with finding C, CC, BC, EVC, and PR. It is more challenging to perform graph-level analysis without any heuristics than node-level analysis. As there could be multiple combinations of nodes in G' depending on various possibilities of node-level parameters and the impact of considering the wrong candidate set for HT could be high, we do pairwise comparison of C, CC, and BC of nodes before the information can be passed down to calculating other properties and confirming node status (Trigger or Payload).  

\begin{figure}
\centering
\includegraphics[width=0.25\textwidth]{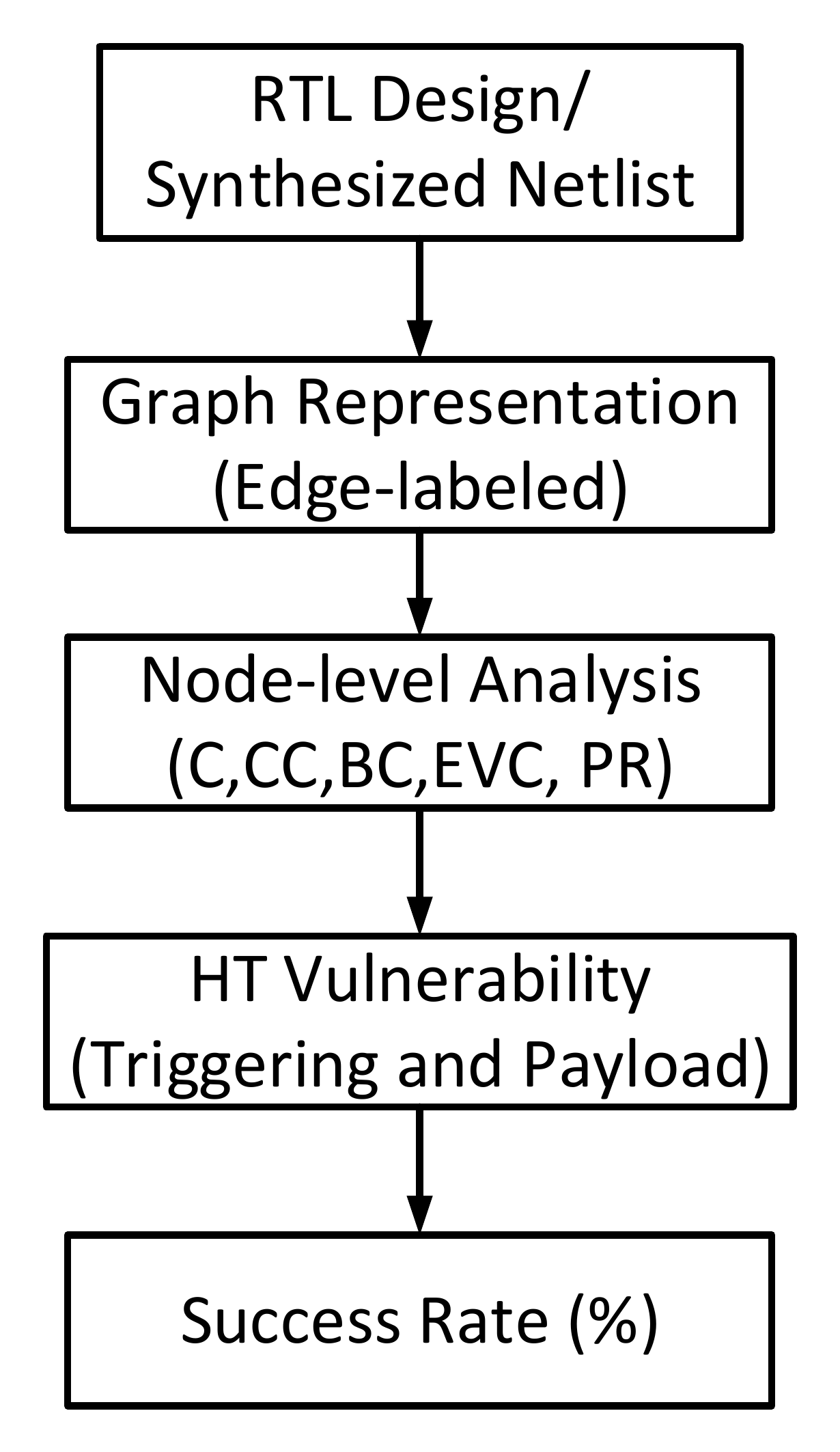}
\caption{Bottom-up analysis framework for HT vulnerability analysis and localization}
\label{fig:framework}
\end{figure}

\begin{figure*}[h]
\centering
\resizebox{\textwidth}{!}{
\begin{tabular}{c c}
\includegraphics[width=\textwidth]{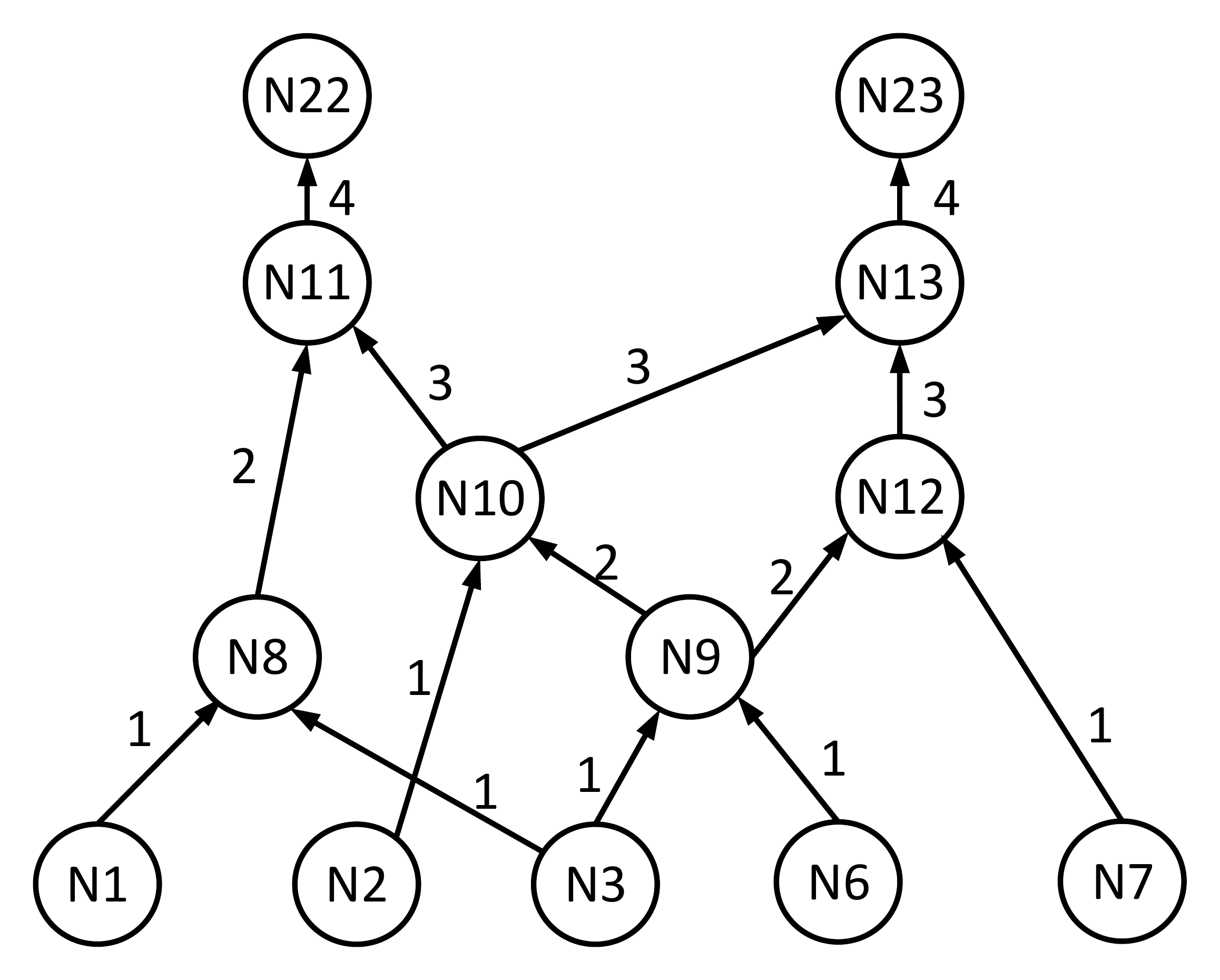} & 
\includegraphics[width=\textwidth]{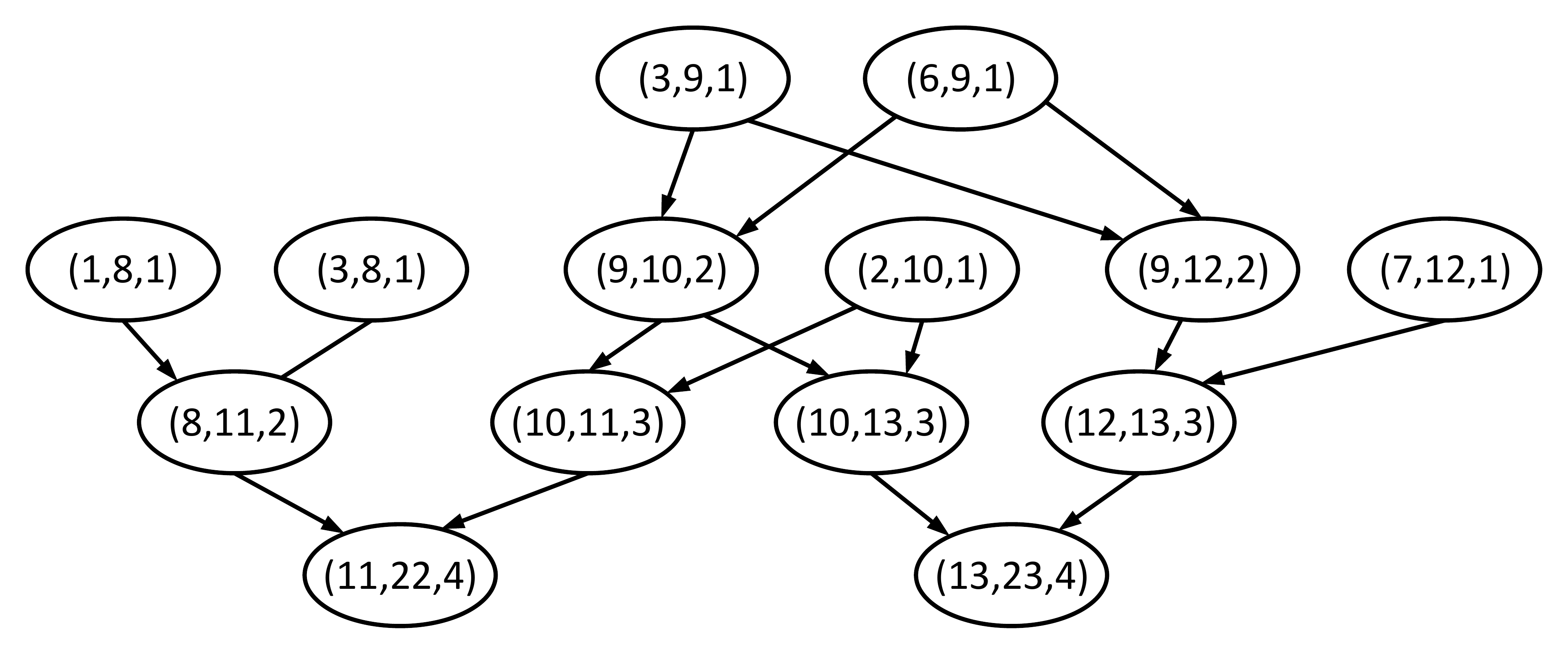} \\ 
\textbf{\LARGE (a)} & \textbf{\LARGE(b)} \\
\end{tabular}}
\caption{(a) Node-labeled c17 circuit and (b) edge-labeled c17 circuit. Both diagrams are HT-free.}
\label{fig:c17_HT_free}
\vspace*{-2ex}
\end{figure*}

With node level data, we can further verify two sub-graph properties (K$_{c}$ and D). In the top-down approach, we start with the graph-based approach (coarse grain technique). Similar to bottom-up, we only consider internal nodes and nets, excluding primary input (P.I.) and primary output (P.O.). The nodes are topologically sorted according to its degree, which are further classified into clusters of a network according to k$_{c}$. We compute the density (D) of all clusters and rank them from lowest to highest. Based on user preference, we can perform further analysis of the nodes within the cluster ranked lowest. Analytically, both approaches should locate the same rare nets responsible for HT. In the rest of the paper, we implement the bottom-up approach. A comprehensive framework for bottom-up analysis is shown in Fig. \ref{fig:framework}.  One can also adopt a top-down approach; however, the success in HT detection would depend on initial heuristics. 

\begin{figure*}[h]
\centering
\resizebox{\textwidth}{!}{
\begin{tabular}{c c}
\includegraphics[width=\textwidth]{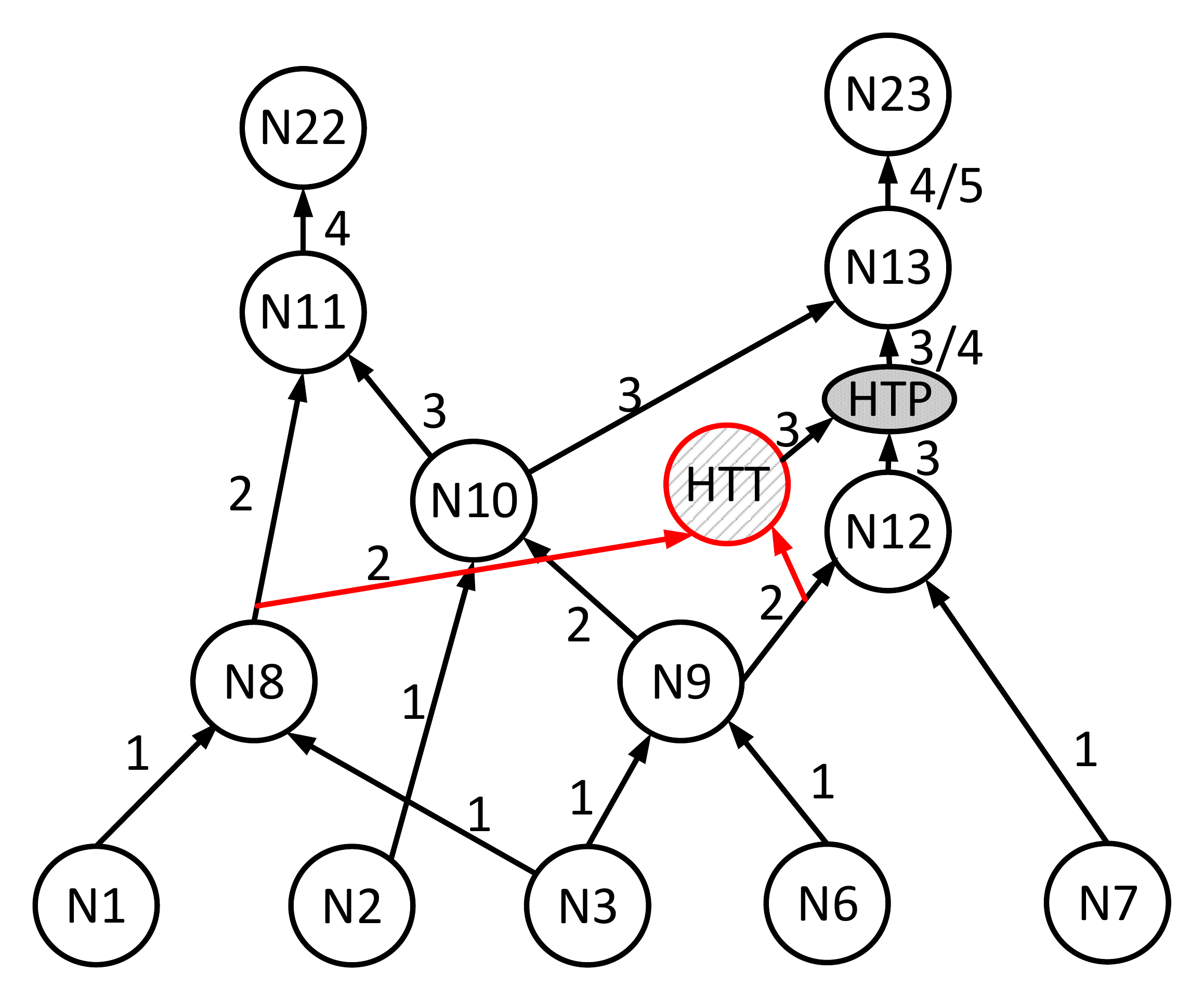} & 
\includegraphics[width=\textwidth]{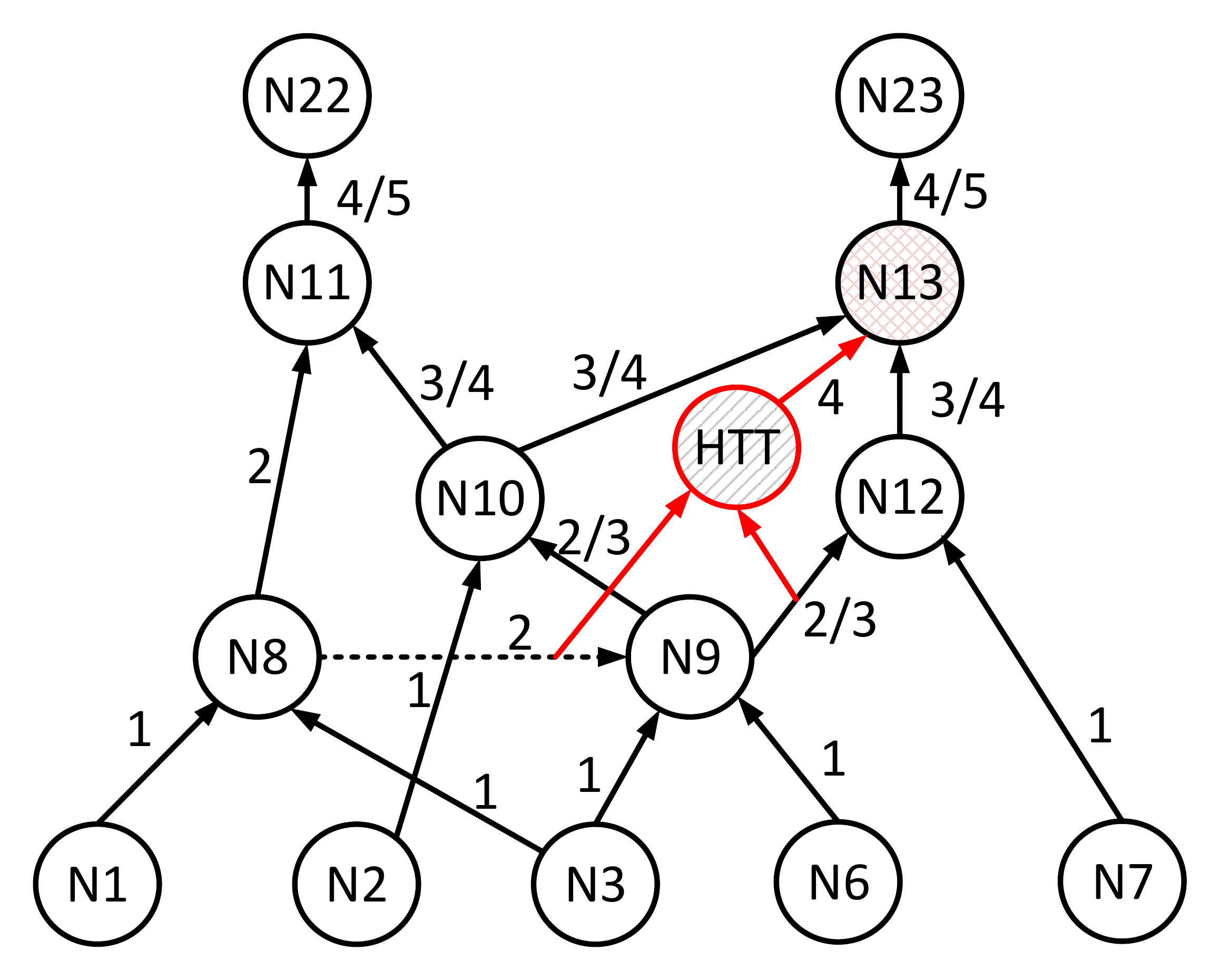} \\ 
\textbf{\LARGE (a)} & \textbf{\LARGE(b)} \\
\end{tabular}}
\caption{Illustration of HT triggering and payload signals for (a) explicit payload and (b) implicit payload.}
\label{fig:c17_HT_payload}
\vspace*{-2ex}
\end{figure*}

Consider a generic c17 circuit (node-labeled) from ISCAS85 benchmark and its corresponding edge-labeled circuit in Fig. \ref{fig:c17_HT_free}.  The design has 5 P.I. and 2 P.O. in the node-labeled graph and edge-labeled ones has 14 nodes which are edges of node-labeled ones. Each edge in the node-labeled graph is labeled incrementally. In the edge-labeled graph, the third parameter associated with each node denotes the label of that node. We remove nodes N1, N2, N3, N6, N7, N22, and N23 from further consideration.

\begin{table}[h]
\begin{center}
\caption{Node attributes for possible HT localization}
\label{tab:signalstat}
\begin{tabular}{|l|c|c|c|c|c|c|c|c|c|c|}
\hline
Node  & C &  CC & BC & EVC & PR \\  \hline

(12,13) & 0.2 & 0.2 & 0 & 0.98 & 0.24 \\ \hline

(9,12) & 0.2 & 0 & 0 & 1.96E-09 & 0.12\\ \hline

(8,11) & 0 & 0 & 0 & 1.16E-019 & 0.12  \\ \hline

(9,10) &  0.4 & 0 & 0 & -2.68E-10 & 0.12 \\ \hline

(10,11) & 0.2 & 0.2 & 0 & -0.13 & 0.18 \\ \hline

(10,13) & 0.2 & 0.2  & 0 & -0.13 & 0.18 \\

\hline
\end{tabular}
\end{center}
\vspace*{-3ex}
\end{table}

In the rest of the section, we would refer to edge-labeled graph representation for analysis. The attributes of the nodes are shown in Table \ref{tab:signalstat}. We can see the five nodes have a centrality value of 0.2 or less, which is also calculated for all nodes in closeness centrality. Then, we  discard node (9,10) from further consideration as it is reachable from the rest of the nodes (C=0.4). As higher CC indicates maximum closeness, we sort the CC column and find that the nodes (9,12) and (8,11) both have minimum  CC as well as moderate closeness (C=0.2). However, for BC, we see the value of each node is `0'. This is because  no shortest path exists between two nodes that would cross any other node in the table.  Hence, the nodes (9,12) and (8,11) can act as HT triggering signals.

On the contrary, EVC and PR are closely related as both can approximate the structure of incoming edge(s). The node having the largest EVC and PR value should be chosen as possible HT payload signal. Hence, the node (12,13) could be used as a payload.

Based on the above relations between nodes, we can classify the possible HT payload into two categories: (a) explicit payload and (b) implicit payload. Explicit payload exhibits a change in regular path delay, whereas implicit payload can circumvent the path delay based fingerprint (i.e., no change in path delay). The impact of both payload is shown in Fig. \ref{fig:c17_HT_payload}. The label (x/y) indicates the change in labeling where the previous label was x, and the new label is y.

During explicit payload, we use existing nets (8,11) and (9,12) to act as HT triggering signal (HTT), and the edge between N12 and N13 is being modified to work as HT payload (HTP) due to the inclusion of output of HTT. In comparison to Fig. \ref{fig:c17_HT_free}(a), the labeling of two edges (N12 $\rightarrow$ N13 and N13 $\rightarrow$ N23)  are modified, which denotes the delay of critical path has been changed.  

For implicit payload, we add an edge between nodes (N8 and N9), which would act as input to HTT with other input being the edge between nodes (N9 and N12). However, we have four choices to create additional nodes between (8,11) and (9,12) who can act as HTT, as shown in Table \ref{tab:nodechoice}. Among four possibilities, we choose node (8,9) as an additional HTT signal. The output of HTT directly modifies the output of N13. If N13 is being replaced with three input NAND, the output will become `0' when all inputs are HIGH (`1') i.e., which is a rare condition (1 out of 8 combinations). Hence, the output of HTT should be `1' to change in the output of N13. Unlike explicit payload, there is no additional gate to act as HTP, and this also ensures the delay of paths towards P.O. is being modified simultaneously.

\begin{table}[h]
\begin{center}
\caption{Characteristics of new node towards implicit HT payload}
\label{tab:nodechoice}
\begin{tabular}{|l|c|}
\hline
Node  & Reasons for choice or not \\  \hline

(8,9) & Impact both outputs' paths without any delay difference  \\ \hline

(8,12) & Impact one output and present path anomaly \\ \hline

(11,12) & Create a back-edge \\ \hline

(9,11) &  Reachable by visiting node N10  \\ \hline

\end{tabular}
\end{center}
\end{table}

\subsection{Algorithm for combinational HT localization}

We propose an algorithm as shown above to  locate nodes that can act as combinational HT. The design is considered to be represented as DAG and edge-labeled. We describe major procedures in the algorithm as follows:

\begin{itemize}
    \item Node extraction: For all nodes in the graph, we mark the P.I., and P.O. and remove them from further consideration.
    \item Triggering nodes location: For the rest of the nodes, we calculate C, CC, and BC value. We sort the node in ascending order based on the value of C. The node(s) in the higher position(s) have larger C value and are marked as influential node(s). Such nodes have a higher fanin and fanout. Hence, adding a new node (edge) to these nodes would have a significant impact on design parameters which the attacker always wants to avoid. Therefore, we remove the influential nodes. For the remaining nodes in the list, we calculate an objective function (F$_{i}$). The nodes showing the lowest F are given priority to be considered as HT triggering signals. 
    \item Payload node location: Once HT signals are known, we sort the nodes (ascending order) according to BC value. The node on the bottom of the sorted list is marked as payload node. We choose this node as it should satisfy the property:
    \textit{it would act as a bridge (hence reachable) if there exist any shortest paths between HT triggering nodes.} Furthermore, the payload node should also exhibit higher EVC and PR value. Higher EVC value confirms that the node is closest to the most influential nodes in the graph, which would manifest them HT triggering. Similarly, depending on damping factor, PR establishes that HTT signals would create new edge the node having highest PR.
\end{itemize}

\begin{equation}
 F_{i} = W_{C}* \frac{C_{i}}{C_{max}} + W_{CC}* \frac{CC_{i}}{CC_{max}}
\end{equation}

\begin{algorithm}[H]
 \KwIn{G'=Edge-labeled combinational design}
 \KwOut{Nodes for HT localization}
 \begin{flushleft}
 G''=Remove the nodes connecting primary inputs and primary outputs  \\
 \end{flushleft}
 \tcc{HT Triggering Nets}
 \For {node in G''} 
 { 
    {\begin{flushleft}
 Calculate C,CC, and BC value \; \\
 Remove the node(s) having highest C and CC value but lowest BC value \; \\
 \end{flushleft}}
 \For {remaining node in G''} 
 {
 Find importance of node by a objective function
 }
 Select top 3-4 nodes as possible HT triggering signals
 }
 \tcc{HT payload}
  \For {node in G''} 
  {
  Calculate EVC and PR value \; \\
 The node showing highest BC, EVC, and PR value is possible HT payload signal \; \\ 
  }
  
Return nodes with label (trigger or payload)
 
\caption{Sorted rare nets set for HT localization}
\label{algorare}
\end{algorithm}



\section{Experimental Results}
\label{sec:result}

We evaluate the socio-network analysis of HT localization using {\textit{NetworkX}} \cite{hagberg2008exploring}. We tested the technique on combinational HTs for ISCAS85 benchmark \cite{8342270}. As mentioned in Section \ref{sec:def}, each parameter influences the HT localization. For each design (HT-free), we perform the same experiment using different configuration parameters (e.g., change in damping value of PR) and we measure the success rate (\%) for all HT types under a particular design. For each trial, we use all HT instances of  design to calculate the success rate. We use a total of ten snapshots of socio-network parameters. We quantify the success in detection using three parameters: 
\begin{itemize}
    \item True Positive (TP): The technique correctly detects the nets responsible for HTT and HTP.
    \item False Positive (FP): The technique detects the nets as HTT and HTP incorrectly in HT infected design.
    \item False Negative (FN): The technique fails to detect nets appearing as HTT and HTP.
    
\end{itemize}

We assume the HT triggering signals would contain at least four nets whereas, for HT payload, it is only one signal. We did not perform any additional experiment such as controllability and/or observability, gate-level simulation to complement our approach. We also did not set transition threshold to refer the net as triggering. The HT infected designs do not contain any implicit payload signal, and we always observe a change in the path delay. Hence, we only presented the success rate for explicit payload. 

Table \ref{tab:result} shows the success rate for four combinational designs. On average, we can detect HTT for 96.25\% time and HTP for 98.5\% time. Given that, there can be multiple HTP signals, the FP and FN rate could be higher. For example, in many HT instances of c6288, there is more than payload signals. Similarly, around 10-15 instances create cycle when combining nets to trigger HT. As our assumption is acyclic graph representation to the algorithm, the method can not detect this HTs. We did not experience any memory bottleneck due to large size of the design. On average, 70\% time is spent to pre-characterize the nets, and the rest goes on checking HT database.

\begin{table}
\begin{center}
\caption{Performance of socio-network parameters in HT detection}
\label{tab:result}
\resizebox{\columnwidth}{!}{
\begin{tabular}{|c| c| c| c| c| c|c|c|c|c|}
\hline
Design & HT instances  & \multicolumn{3}{c|}{HTT}  & \multicolumn{3}{c|}{HTP} \\
\cline{3-8}
& & TP & FP & FN & TP & FP & FN \\ \midrule
 c2670 & 100 & 97\% & 1\% & 2\% & 100\% & 0\% & 0\% \\ 
 c3540 & 100 & 95\% & 3\% & 2\% & 99\% & 0\% & 1\% \\ 
 c5315 & 110 & 95\% & 1\% & 4\% & 98\% & 1\% & 1\% \\
 c6288 & 110 & 98\% & 2\% & 0\% & 97\% & 2\% & 1\% \\ \midrule
 Average & & 96.25\% & 1.75\% & 2\% & 98.5\% & 0.75\% & 0.75\% \\
 \midrule

\end{tabular}
}
\end{center}
\vspace{-5ex}
\end{table}

\section{Conclusion and Future Work}
\label{sec:conclude}

We present a framework to detect diversity-agnostic HT based on social network parameters. With only knowledge of the design, a designer can find possible HT locations easily without any expensive simulation or side-channel analysis. We also present an algorithm to rank the parameters with  lower time complexity. We describe the feasibility of bottom-up analysis approach. However, given enough run-time and good heuristics, top-down analysis is also possible. Future work includes the same parameters review for sequential HT and at a higher abstraction level (e.g., algorithmic level).

\bibliographystyle{unsrt}
\scriptsize{
\bibliography{HT.bib,pba_ht.bib}
}
\end{document}